\documentclass[aps,prb,preprint,showpacs,groupedaddress]{revtex4} 

\usepackage{graphicx}

\begin{document}

\title{
Spin-dependent correlation in two-dimensional electron liquids
at arbitrary degeneracy and spin-polarization: CHNC approach
}

\author{Nguyen Quoc Khanh}
\affiliation{ Department of Theoretical Physics, National
University in Ho Chi Minh City, 227-Nguyen Van Cu Str., District
5, Ho Chi Minh City, Vietnam }

\author{Hiroo Totsuji}
\email[]{totsuji@elec.okayama-u.ac.jp}
\affiliation{
Department of Electrical and Electronic Engineering,
Okayama University,
Tsushimanaka 3-1-1, Okayama 700-8530, Japan
}

\date{\today}

\begin{abstract}
We apply the classical mapping technique developed recently by
Dharma-wardana and Perrot for a study of the uniform
two-dimensional electron system at arbitrary degeneracy and
spin-polarization. Pair distribution functions, structure factors,
the Helmhotz free energy, and the compressibility are calculated
for a wide range of parameters. It is shown that at low
temperatures $T/ T_F <0.1$, $T_F$ being the Fermi temperature, our
results almost reduce to those of zero-temperature analyses. In the
region $T/ T_F \ge 1$, the finite temperature effects become
considerable at high densities for all spin-polarizations.
We find that, in our approximation without bridge functions,
the finite temperature electron
system in two dimensions remains to be paramagnetic fluid
until the Wigner crystallization density. Our results are
compared with those of three-dimensional system and indicated are
the similarities in temperature, spin-polarization, and density
dependencies of many physical properties.
\end{abstract}


\maketitle

\section{Introduction}

Two-dimensional electron systems
have been investigated by a number of investigators.
Their importance seems to have recently increased
due to possible applications to spintronics
for electronic devices.
The properties of two-dimensional electron systems
are determined by the Coulomb interaction between electrons
and
depend crucially on the density, temperature and spin-polarization.
In principle,
exact results can be obtained by the numerical experiments
and
are available to some extent at zero temperature.
At finite temperatures,
however,
reliable numerical experiments are still awaited
and
the importance of theoretical approach is not reduced.

Recently, Dharma-wardana and Perrot have developed a method called
classical-map hypernetted-chain method
(CHNC)\cite{DP00,PD01,PD00b,DP03}.
They determine the quantum
temperature $T_q$ so that the classical system at the temperature
$T_q$ has the same correlation energy as the quantum system at
$T=0$ and assume that the properties of the system at the finite
temperature $T$ is given by those of classical system at the temperature
\begin{equation}\label{T_cf}
T_{cf}=(T_q^2+T^2)^{1/2}.
\end{equation}
One can include the
higher-order cluster interactions in classical fluids applying the
modified HNC method\cite{RA79} and taking the bridge function into account.

Various properties of two-dimensional electronic systems at
zero-temperature have been examined by Bulutay and
Tanatar\cite{BT02} based on the CHNC method without bridge
corrections. We here extend the analysis to finite temperature
two-dimensional electronic systems with arbitrary
spin-polarization. Confirming that our results reduce to those of
Ref. 6 at low temperatures such that $T/T_F \le 0.1$, we analyze
the finite temperature effects for $T/ T_F \ge 0.5$, $T_F$ being
the Fermi temperature. We observe that the effects are substantial
at high densities irrespective of spin-polarization. It is shown
that, within our analyses without bridge functions, the
two-dimensional electron system remains paramagnetic until the
Wigner lattice formation. In Ref. 4, the existence of a
ferromagnetic phase before the formation of Wigner lattice is
predicted and, in such a domain, one needs careful treatment
including the bridge function. The purpose of this paper is to
analyze fundamental quantities in a wide domain of density,
temperature, and spin-polarization where one may neglect the
bridge function as in Ref. 6. In what follows, we adopt the atomic
units and take $k_B=1$.

\section{CHNC method applied to two-dimensional electron system}

We consider a two-dimensional fluid of electrons at the temperature $T$
containing two spin species
of surface densities $n_1$ and $n_2$ $(n_1 \ge n_2)$
with the total density $n=n_1+n_2$ and spin-polarization $\zeta= (n_1-n_2)/n $.
We determine the Pauli potential for parallel-spin electrons $P_{ii}(r)$
by the relation
\begin{equation}
g_{ii}^0(r) \,=\,\exp[-\beta P_{ii}(r) + h_{ii}^0(r) -
c_{ii}^0(r)].
\end{equation}
Here $\beta$ is determined by the temperature $T_{cf}$ by $\beta=1/T_{cf}$,
$ g_{ii}^0(r)= h_{ii}^0(r)+1$ is the ideal gas pair distribution function
for the component $i$,
\begin{equation}
h_{ii}^0(r)\,=\,-\frac{1}{n_{i}^2}\sum_{{\bf k}_1,{\bf
k}_2}\,n(k_1)n(k_2)\, \exp[i({\bf k}_1-{\bf k}_2)\cdot{\bf
r}]\,=\,-[f_i(r)]^2,
\end{equation}
and
$c^0_{ii}(r)$ is the direct correlation function corresponding to $ h_{ii}^0(r)$.
In (3),
$n(k)$ is the Fermi occupation number at temperature $T$.
At $T = 0$,
$f_i(r)=2J_1(k^i_F r)/ k^i_F r$,
where $J_1(x)$ is the Bessel function and $k^i_F$ is the Fermi wave number
of species $i$.

The pair distribution functions (PDF's) and the direct correlation functions
are related via
\begin{equation}\label{pdf}
g_{ij}(r)\,=\,\exp\,[-\beta\,\phi_{ij}(r)+h_{ij}(r)
-c_{ij}(r)+B_{ij}(r)].
\end{equation}
Here
$\phi_{ij}(r)$ is the pair potential between the species $i$ and $j$ defined by
\begin{equation}
\phi_{ij}(r)\,=\,P_{ii}(r)\delta_{ij}+V_{\rm Cou}(r),
\end{equation}
and $h_{ij}(r)=g_{ij}(r)-1$, $c_{ij}(r)$, and $B_{ij}(r)$ are the
pair correlation, direct correlation and the bridge function,
respectively. In the pair potential, $ V_{\rm Cou}(r)$ includes
the diffraction effect as
\begin{equation}
V_{\rm Cou}(r)\,=\,\frac{1}{r}[1-e^{-rk_{\rm th}}],
\end{equation}
where $k_{th}=(2\pi m^*T_{cf})^{1/2}$, $m^*=1/2$ is the reduced
mass of the scattering electron pair \cite{HM81}. We also have the
Ornstein-Zernike relations between the pair correlation functions
and the direct correlation functions:
\begin{equation}\label{OZ}
h_{ij}(r)\,=\,c_{ij}(r)
+\sum_s\,n_s\,\int\,d{\bf r}^\prime\,h_{is}(|{\bf r}-{\bf r}^\prime |)
c_{sj}(r^\prime).
\end{equation}
The equations (\ref{pdf}) and (\ref{OZ}) are exact but are not
closed because of the existence of the unknown bridge functions
$B_{ij}(r)$. In this paper, however, we neglect the bridge
functions as in Ref. 6 and solve the set of equations (\ref{pdf})
and (\ref{OZ}). We have analyzed\cite{KT03SSC} the case of $r_s=5$
with the bridge function employing the solution of the
Percus-Yevick equation for hard disks with somewhat different
formula for $T_q$. The result indicates that the effect of the
neglect of bridge function increases with the temperature and the
free energy due to exchange-correlation is overestimated by at
most 5\% at $T/T_F=5$. The results for fundamental quantities
obtained without bridge functions will therefore be still useful
in a wide range of density, temperature, and spin-polarization
where the effect of the neglect of the bridge function is not so
subtle.

In the CHNC method, the temperature of the classical fluid $
T_{cf}$ is determined by (\ref{T_cf}). For the quantum temperature
$T_q$, we adopt the one given by Ref. 6,
\begin{equation}
T_q\,=\,\frac{1+ar_s}{b+cr_s}\,\frac{n_1T_{F1}+n_2T_{F2}}{n}
\end{equation}
where $r_s=(\pi\,n)^{-1/2}$ with $a = 1.470342$, $b = 6.099404$, and $c = 0.476465$.

The exchange-correlation part of the Helmholtz free energy is calculated by the integration over the coupling $\lambda$ as
\begin{equation}
\frac{F_{xc}}{n}\,=\,\pi\,n\,\int_0^1\,d\lambda\,\int\,dr\,[g(\lambda; r)-1]
\end{equation}
from the spin-averaged PDF given by
\begin{equation}
g(r)\,=\,\frac{1}{4}[(1+\zeta)^2\,g_{11}(r)+2(1-\zeta^2)g_{12}(r)
+(1-\zeta)^2g_{22}(r)].
\end{equation}
The total Helmholtz free energy $F_{tot}$ is obtained as
\cite{DP03}
\begin{equation}
F_{tot}\,=\,F_0+F_{xc},
\end{equation}
where $F_0= F_0^1+ F_0^2$ is the free energy of the ideal electron gas
given by
\begin{equation}
F_0^i\,=\,n_i\mu_0^i - E_0^i,
\end{equation}
\begin{equation}
\mu_0^i\,=\,T\,\ln[e^{E_F^i/T}-1],
\end{equation}
\begin{equation}
E_F^i\,=\,(1\pm \zeta)\,\pi\,n,
\end{equation}
and
\begin{equation}
E_0^i\,=\,\frac{T^2}{\pi}\,\int_0^{\infty}\,\frac{dx\,x}{\exp[x-\mu_0^i/T]+1}.
\end{equation}

We calculate the exchange part of the Helmholtz free enegy
$F_x\,=\,F_x^1+F_x^2$ by the ideal gas values as
\begin{equation}
\frac{F_x^1}{n}\,=\,\frac{1}{4}\,\pi\,n(1+\zeta)^2\, \int\,dr\,
[g_{11}^0(r)-1],
\end{equation}
\begin{equation}
\frac{F_x^2}{n}\,=\,\frac{1}{4}\,\pi\,n(1-\zeta)^2\, \int\,dr\,
[g_{22}^0(r)-1],
\end{equation}
and define the correlation part by $F_c=F_{xc}-F_x$.
At $T = 0$
we have
\begin{equation}
{F_x \over n}={E_x \over n}
=\,-\frac{2\sqrt{2}}{3\pi\,r_s}[(1+\zeta)^{3/2}+(1-\zeta)^{3/2}],
\end{equation}
\begin{equation}
{E_0\over n}\,={E_0^1+E_0^2\over n}=\,\frac{1+\zeta^2}{2r_s^2}.
\end{equation}

\section{Results and discussions}

We have solved the coupled equations (\ref{pdf}) and (\ref{OZ})
using the Hankel transform as in Refs. 6 and 9 and neglecting the
bridge functions for a wide range of density, spin-polarization
and temperature, $0 \le r_s \le 30$, $0 \le \zeta \le 1$, and $0
\le t\equiv\,T/T_F \le 5$. Here we summarize the results.

\subsection{Pair distribution function and structure factor}

The spin-averaged PDF $g(r)$ for $r_s = 1$ and $T/T_F =
0, 1$ in two cases $\zeta= 0, 1$ is shown in Fig.1. Our results
for $T/T_F =0$ are in good agreement with those given in Ref. 6.
We observe that the finite temperature effect considerably reduces
the exchange-correlation hole around electrons in both cases of
$\zeta=0$ and $\zeta=1$ and we may estimate the behavior of the
pair distribution function in the case of arbitrary spin
polarization from these results.

In Fig. 2,
we show the spin-averaged structure factor defined by
\begin{equation}
S(q)\,=\,1+n\,\int\,d{\bf r}\,[g(r)-1]\,e^{i{\bf q}\cdot{\bf r}}
\end{equation}
for $r_s = 1$, $T/T_F = 0, 1$, and $\zeta = 0, 1$. At $T/T_F = 0$,
the structure factor is consistent with the one given in Ref. 6.
We observe the finite temperature effect for the cases of
spin-polarization $\zeta=0, 1$.

\subsection{Total free energy}

We have calculated the total Helmholtz free energy for finite
temperatures $0 \le T/T_F \le 5$ and spin-polarizations $0 \le
\zeta \le 1$. The results are shown as a function of the density
parameter $r_s$ and spin-polarization $\zeta$ in Figs. 3 and 4,
respectively.

At low temperatures $T/ T_F \le 0.1$, our results reduce to those
of Ref. 6. When $T/ T_F \ge 1$, the finite temperature effects
become considerable at high densities for all cases of
spin-polarization. Our results indicate that the two-dimensional
electron system remains to be in the paramagnetic fluid phase
until the Wigner crystallization densities even at high
temperatures.

The dashed-dotted line in Fig. 4 represents the total free energy
for $T/T_F=1$ and $r_s = 1$.
It is seen from the figure that
the spin-polarization effect is more pronounced at higher temperatures and densities.

\subsection{Exchange-correlation free energy}

To compare the electron correlation effects in two and three dimensions,
we plot the exchange-correlation and correlation free energies
as a function of temperature in Figs. 5 and 6.
We find that
our results of the temperature dependence of the free energy in two dimensions
are similar to those in three dimensions\cite{PD00b}.

To check the validity of the CHNC scheme used in this paper, we
compare our exchange-correlation energy with Monte Carlo (MC) and
Singwi-Tosi-Land-Sj\"{o}lander scheme (STLS) results given in
Refs. 10, 11, and 12. The exchange-correlation energies computed by
different schemes are listed in Table I. We observe that the
difference between our results and STLS results is considerable
(about 10\%) only at high temperatures $(T>T_F)$. This may be
because we have used the expression for quantum temperature $T_q$
fitted to MC data at $T=0$. To obtain better agreements with STLS
results for $T>T_F$ we may need to include the bridge term and use
the results of finite temperature STLS scheme given in Ref. 12 for
fitting procedure.

As for the exchange-correlation free energy at high densities,
the agreement with MC and other values
can be improved by interpolating $T_q$ more accurately.
For example,
the set
$a=24.936526$, $b=87.221405$, and $c=8.455547$
gives better fitting for $T_q$ as shown in Fig. 7
and better agreement as listed on the last line in Table I.

\begin{table}
\caption{Exchange-correlation energy  per  electron  (in  a.u.) of
unpolarized  2DEG, $-F_{xc}$. For comparison, the MC results of
Tanatar and Ceperley\cite{TC89} and the STLS results of
Jonson\cite{MJ76} and Schweng and B\"{o}hm\cite{SB94} are given
(denoted by a, b, and c, respectively). The results  for  fully
polarized  2DEG  are shown  in  brackets. The last line is the
values based on improved interpolation for $T_q$.
}
\begin{ruledtabular}
\begin{tabular}{c|ccccccccccc}
$T/T_F$  &  0.0 &  $0.0^a$  &  $0.0^b$  &  0.1 &  $0.1^c$ &  0.5  &  $0.5^c$  &
1.0  & $1.0^c$ & 5.0 &  $5.0^c$ \\ \hline
$r_s=1$ & 0.688  &   0.708  &   0.705  &   0.686  &    0.708 &     0.641  &
 0.685   &    0.583   &   0.633    &  0.359  &  0.399\\
5   &   0.167  &   0.168    &   & 0.167   &   0.167   &   0.163  &
     0.167    &   0.158   &   0.163 &     0.121  &  0.127 \\
 & (0.181) & (0.180) \\
10  &   0.089   &  0.089   &      &    0.089  &    0.089   &   0.088  &
      0.088   &    0.086   &   0.073  &    0.071  &  0.063 \\
 &  (0.092)  & (0.091) \\ \hline
$r_s=1$ & 0.696  &   0.708  &   0.705  &   0.693  &    0.708 &     0.646  &
 0.685   &    0.585   &   0.633    &  0.359  &  0.399\\
\end{tabular}
\end{ruledtabular}
\end{table}

\subsection{Compressibility}

The compressibility is obtained by density derivatives of the free
energy. In order to check the accuracy of our finite temperature
scheme, we have calculated the compressibility as a function of
density parameter $r_s$ or spin-polarizations $\zeta$ for
different temperatures. We note that at low temperatures our
results reduce to those of Ref. 6. Our results shown in Fig. 8
indicate that the temperature effect on the compressibility is
remarkable only at low densities and high temperatures such that
$T/T_F >0.3$.

The spin-polarization dependence of the compressibility is displayed in Fig. 9.
We observe that
the effect of spin-polarization increases with the temperature and electron density.

\section{Conclusions}

Applying the recently proposed classical-map hypernetted-chain
(CHNC) method to the two-dimensional electron system, we have
calculated the pair distribution  function, structure factor,
Helmhotz free energy, and compressibility at finite temperatures
for a wide range of spin-polarization and density. It is shown
that at low temperatures $T/T_F \le 0.1$, our results are almost
identical to those of Ref. 6 at $T=0$. Our results indicate that
correlation characteristics in 2D electron liquids depend
remarkably on the temperature, spin-polarization, and density,
similarly to the three-dimensional case.

When $T>T_F$,
the finite temperature effects become considerable at high densities
irrespective of spin-polarization
as concretely shown in various quantities in this paper.
We find that
the finite temperature 2D electron system remains
to be in the paramagnetic fluid phase
until the Wigner crystallization density is attained.
Though the validity of the conclusion on the phase around $r_s \sim 30$
might depend on that of CHNC
applied to two-dimensional electron system
and the neglect of bridge functions,
results obtained here for $1< r_s <30$ will be useful
in investigation of the system
whose finite temperature properties are important in applications
but not exactly known.

\begin{acknowledgments}
One of us, Nguyen Quoc Khanh, gratefully acknowledges the
financial support from the Japan Society for the Promotion of
Sciences (JSPS). He would also like to thank members of H.
Totsuji's research group at Okayama University, especially Drs. C.
Totsuji and K. Tsuruta, for warm hospitality and helpful
discussions.
\end{acknowledgments}

\newpage
\begin{figure}
\includegraphics{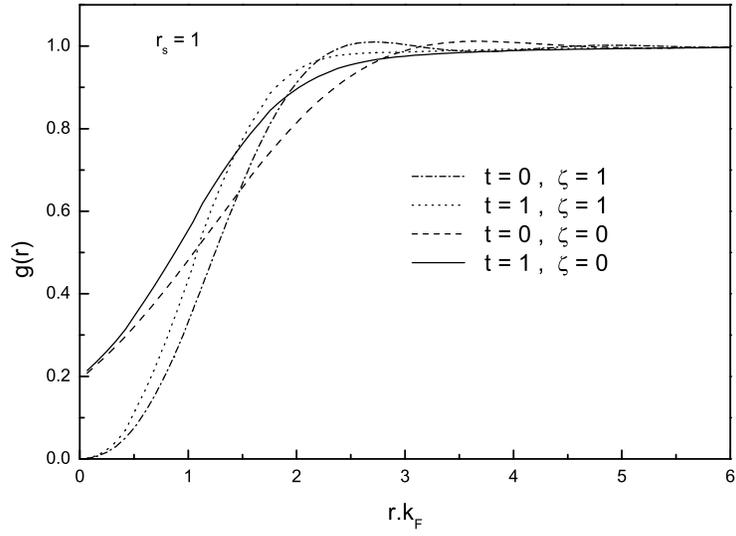}
\caption{Pair distribution function $g(r)$ for $r_s=1$, $t=T/T_F=0,
1$, and $\zeta=0, 1$.\label{pdf-fig}}
\end{figure}

\begin{figure}
\includegraphics{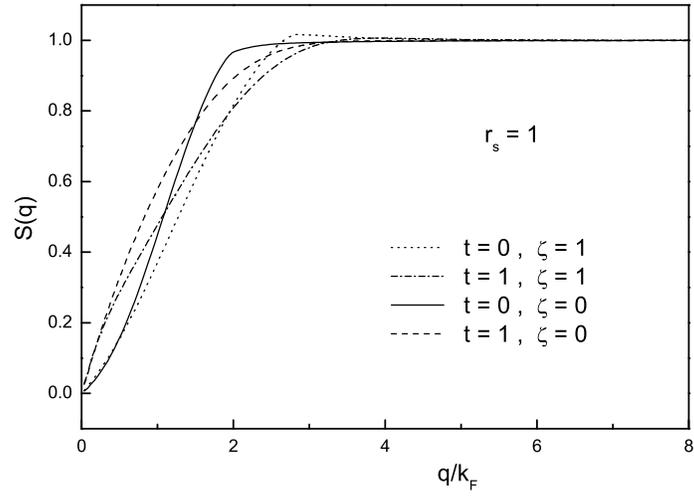}
\caption{Structure factor $S(k)$ for $r_s=1$, $t=T/T_F=0, 1$, and
$\zeta=0, 1$.\label{S(k)}}
\end{figure}
\begin{figure}
\includegraphics{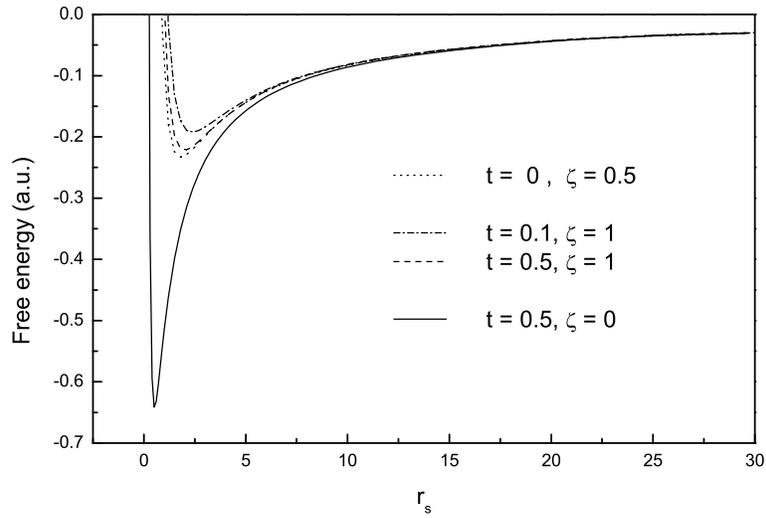}
\caption{Total Helmholtz free energy as a function of $r_s$ for
different temperatures and spin-polarization.\label{free energy vs
rs}}
\end{figure}
\begin{figure}
\includegraphics{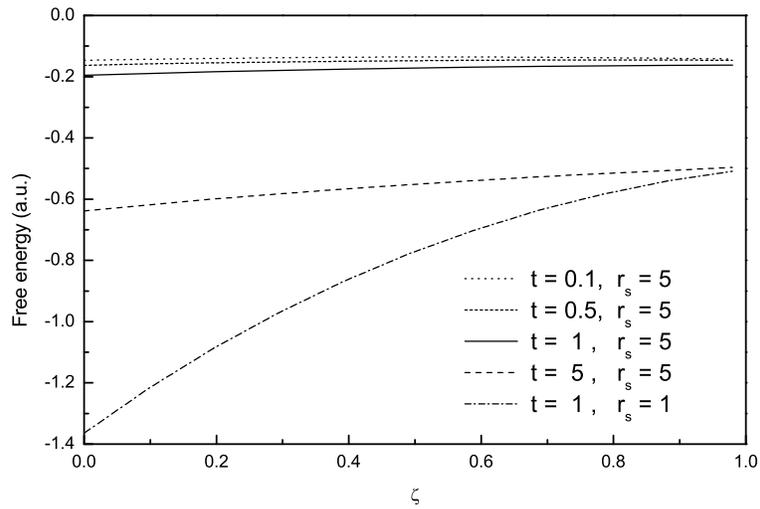}
\caption{Total Helmholtz free energy as a function of
spin-polarization $\zeta$ for different temperatures and
densities.\label{free energy vs zeta}}
\end{figure}
\begin{figure}
\includegraphics{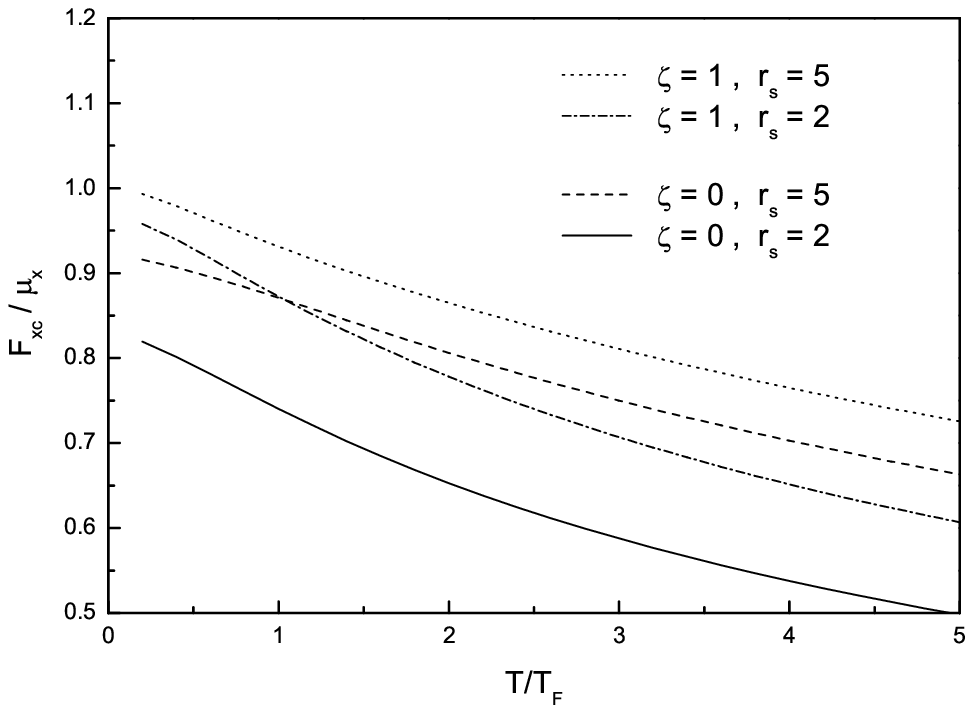}
\caption{Exchange-correlation free energy per electron in units of
$\mu_x=-2k_F/\pi$
 as a function of temperature
for different spin-polarization and
densities.\label{exchange-correlation free energy vs T}}
\end{figure}
\begin{figure}
\includegraphics{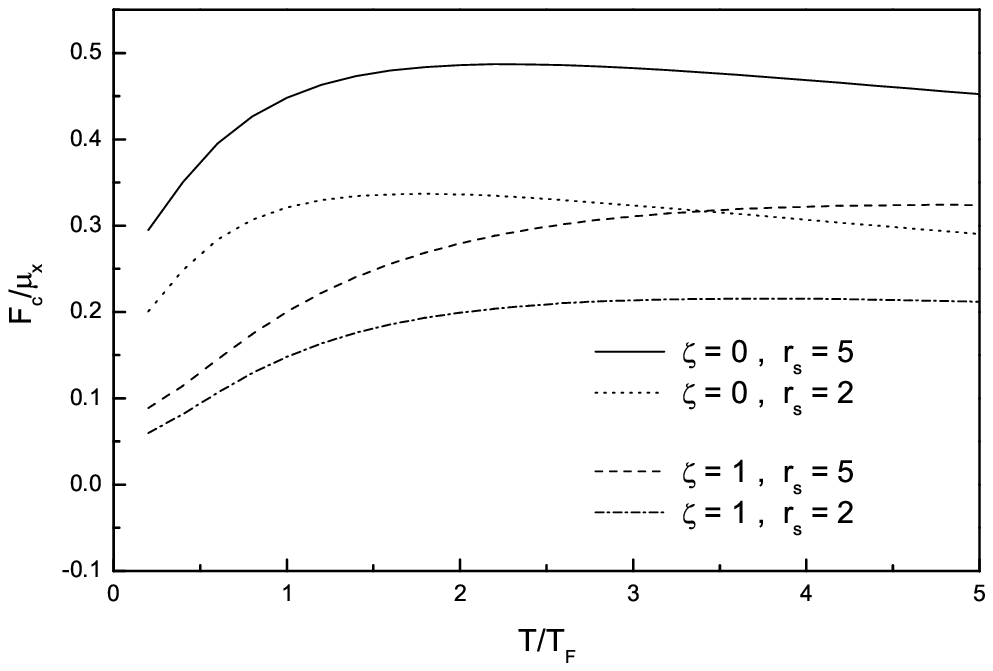}
\caption{Correlation free energy per electron in units of
$\mu_x=-2k_F/\pi$
 as a function of temperature
for different spin-polarization and densities.\label{correlation
free energy vs T}}
\end{figure}
\begin{figure}
\includegraphics{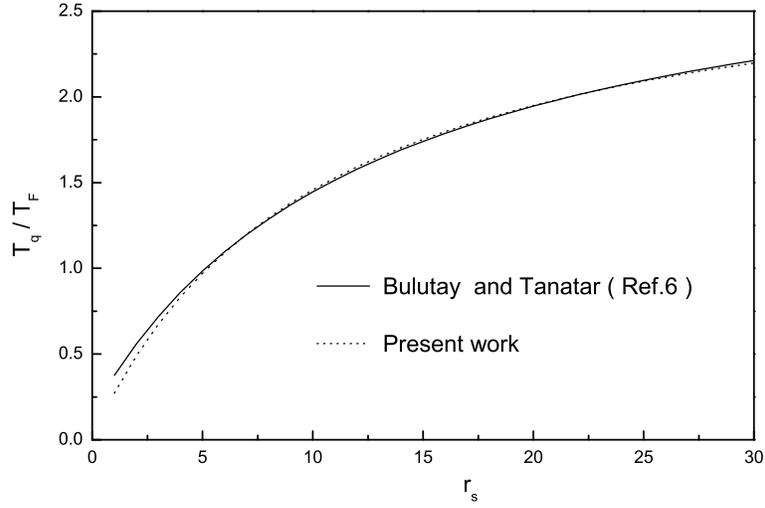}
\caption{Improved fitting for $T_q$ in comparison with the one
given in Ref. 6. \label{fitting}}
\end{figure}
\begin{figure}
\includegraphics{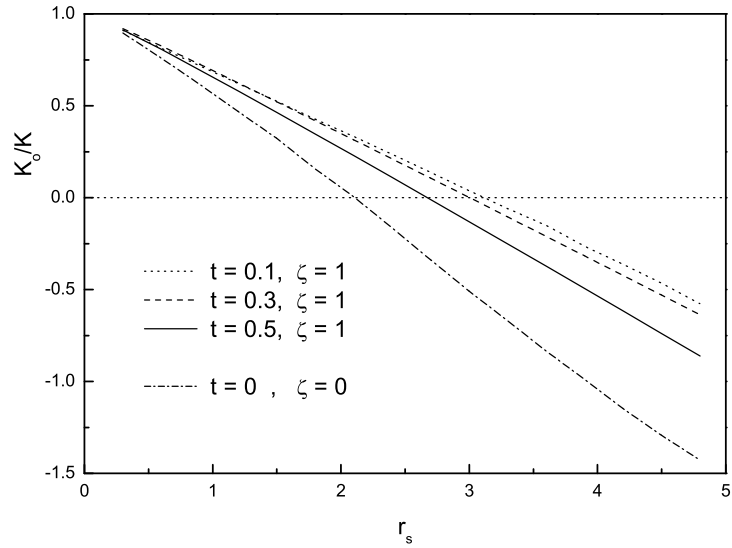}
\caption{Inverse compressibility of unpolarized $(\zeta=0)$ and
fully polarized  $(\zeta=1)$ phases normalized by 2D free Fermion
value $K_0$ for $t=T/T_F=0, 0.1, 0.3, 0.5$. \label{inverse
compressibility}}
\end{figure}
\begin{figure}
\includegraphics{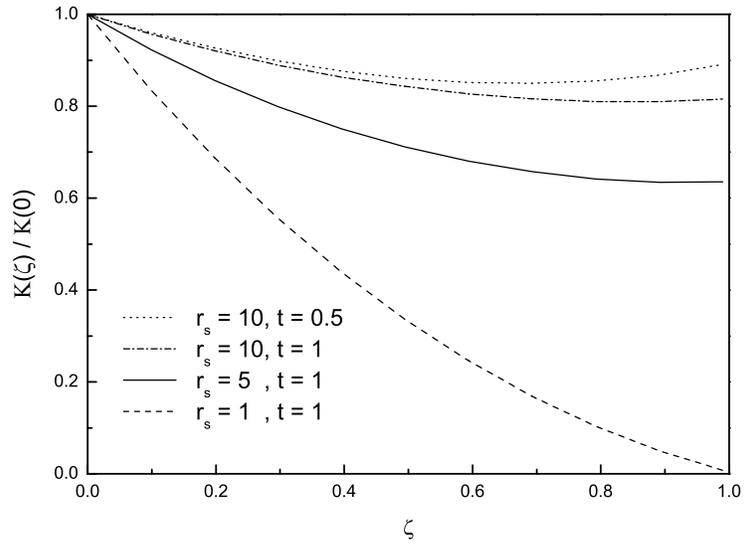}
\caption{Compressibility normalized by paramagnetic value
$(\zeta=0)$ as a function of spin-polarization for $t=T/T_F=0.5, 1$,
and $r_s=1, 5, 10$. \label{compressibility vs zeta}}
\end{figure}

\end{document}